# The Aggregator Model of Spatial Cognition

Robert Worden

UCL Theoretical Neuroscience Group

rpworden@me.com

January 2018, revised October 2020

**Abstract:**

Tracking the positions of objects in local space is a core function of animal brains. We do not yet fully understand how it is done with limited neural resources. The challenges of spatial cognition are discussed under the criteria: (a) scaling of computational costs; (b) feature binding; (c) precise calculation of spatial displacements; (d) fast learning of invariant patterns; and (e) exploiting the strong Bayesian prior of object constancy.

The predominant current models of spatial cognition are Hierarchical Bayesian models of vision, and Deep Neural Nets. These are typically **fully distributed** models, which compute using direct communication links between a set of modular knowledge sources, and no other essential components. This distributed nature leads to difficulties with the criteria (a) - (e).

I discuss an alternative model of spatial cognition, which uses a single central **position aggregator** to store estimated locations of each object or feature and applies constraints on locations in an iterative cycle between the aggregator and the knowledge sources. This model has advantages in addressing the criteria (a) - (e).

If there is an aggregator in mammalian brains, there are reasons to believe that it is in the thalamus. I outline a possible neural realisation of the ensuing aggregator function in the thalamus.

## 1. Introduction

This paper examines spatial cognition - how animals understand the positions and movements of objects in the local space around them, combining sense data of different modalities.

In recent years there has been intense activity addressing spatial cognition - both in understanding how it is done in animal brains, and in the engineering problem of doing it in modern computers, for diverse applications.

The leading models of animal spatial cognition are hierarchical Bayesian models of vision (Lee & Mumford, 2003; Friston 2003; Kersten & Yuille 2003; George & Hawkins 2004; Fei-Fei & Perona 2005; Kokkinos & Yuille 2009), and other hierarchical models such as Rolls (2012). In these models, modular knowledge sources in the cortex cooperate directly with one another, in both a bottom-up and top-down manner, to compute the positions and motion of objects in local space.

While these models have had some successes, huge challenges remain in accounting for observed animal performance with any neural architecture. These challenges are discussed under the following related headings:

a) **Scaling** of the costs of computation and communication, as the number of tracked features and the number of knowledge sources grow
b) **Feature binding** - defining the identity of features, so that different knowledge sources can collaborate to fix the locations of the same feature
c) **Precise calculation** of spatial displacements between features
d) **Rapid learning** of spatial patterns, even when training examples are presented at a variety of locations and orientations.
e) **Exploiting object constancy** - the fact that for most of the time, most things do not move.

In artificial computer vision and spatial perception, the leading current models are based on Convolutional Neural Nets (LeCun, Bengio & Hinton [2015]). These have similarities to hierarchical Bayesian models, in that the successive layers of the neural net correspond to different levels of aggregation of features, in a roughly hierarchical manner; and that the learning units are linked directly to one another. There are both bottom-up and top-down interactions between a large number of learning modules.

In spite of their use of huge (and biologically impossible) amounts of training data, and other benefits of modern computers such as unlimited spatial precision, artificial vision systems have not yet reached animal-like performance, in their ability to track and categorise many types of objects in the presence of visual clutter, occlusion, and so on.

This paper examines the challenges (a) - (e) faced by biological and artificial models. It concludes that these challenges are severe for the biological models, largely because they are **fully distributed** models - in which spatial configurations are computed by the direct collaboration of a set of modular, distributed knowledge sources, with hard-wired direct neural connections between them.

An alternative model is examined - the **aggregator model**, in which the brain stores a single central representation of the position of each feature being tracked, in a structure which is called an aggregator. The aggregator model supports an iterative two-phase algorithm for feature tracking:

- In phase 1, each knowledge source independently applies its own small set of constraints on the positions of a small set of features, to make a local maximum likelihood estimate of their relative positions - and passes the results to the aggregator.
- In phase 2, for each tracked feature, the aggregator combines the position estimates from each knowledge source, to make a pooled maximum likelihood estimate for the position of that feature - and passes the results back to the knowledge sources.

This iteration splits spatial cognition into a number of small independent computations - allowing a high level of parallelism between different knowledge sources, and between different tracked features in the aggregator.

I compare the fully distributed model with the aggregator model, under the headings (a) - (e) above. Because the aggregator model does not require the knowledge sources to be directly connected to one another - but allows them to communicate position information indirectly through the aggregator - it allows information about the locations of features to be shared between knowledge sources using a much smaller number of communication paths. This is more economical than a distributed model, in the brain energy costs of long axons. It can be argued that the number of axons which would be required for a fully distributed architecture is greater than that observed in mammalian brains.

The aggregator architecture has benefits over the fully distributed architecture in the other factors (b) - (e). In order to deliver these benefits, there are demanding requirements on the performance of the aggregator itself. These requirements are well-defined, and appear to be tractable. They are addressed in a later section of the paper, and are the subject of another paper in preparation.

If mammalian spatial cognition uses an aggregator, there is a simple argument that the aggregator must be close to the physical centre of the brain. The aggregator needs to communicate in both directions with knowledge sources distributed around an approximately spherical cortex. The net length of the axons required (and thus cost in brain energy consumption) is minimised by placing the aggregator near the centre of the brain.

This and other facts support a hypothesis that **the aggregator is the thalamus**:

1. The thalamus is prominent and is centrally located in all mammalian brains
2. The thalamus is a gateway to cortex for sense data of nearly all modalities, as is required to use these sense data to locate objects and features in space.
3. Notably, the thalamus is not a gateway for olfactory data. This is because smells are of little use in defining local positions; a smell can come from anywhere, and smell data has poor time resolution.
4. The thalamus has reciprocal connections to many areas of cortex, as is required for two-way communication with diverse knowledge sources.
5. Waking thalamo-cortical rhythms can be identified with the aggregator cycle
6. The thalamus has sufficient size in all dimensions to support a high-resolution storage of three-dimensional position data.

These facts are consistent with the hypothesis that the thalamus acts as an aggregator for three-dimensional position information - and that this is its main function. This model is an instance of the model proposed by Mumford [1991] in which the thalamus acts as a 'blackboard' (Erman et al [1080]) collating information from different cortical knowledge sources. In this case, it collates specifically location information. Other models of thalamic function (e.g. Jones [2007] and Sherman & Guillery [2006]) do not account so well for its important central position in the brain.

I next describe in outline how neural structures in the thalamus might support the precise representation of spatial locations, and the precise calculation of spatial displacements, as is required for the aggregator function.

There is a worthwhile research program to look beyond the current state of the fully distributed models, to consider how they can address the challenges (a) - (e) above - while at the same time examining alternatives such as the aggregator model, which may have a better chance of doing so. In many cases, this need not involve abandoning the distributed models. It would involve adding an aggregator component to them.

To summarise how the model of this paper differs from the most popular models of spatial cognition:

At Marr's (1980) Computational level, the leading models are hierarchical Bayesian models, with sets of Bayesian inference units cooperating in a hierarchy, to infer the locations of objects in space.

At Marr's 1980 Neural Implementation level 3, hierarchical Bayesian models are typically mapped onto brains as follows:

| Level 1 Component | Level 3 Implementation |
|---|---|
| Bayesian inference unit | Cortical module |
| Links between inference units | Cortico-cortical axons |

The first point of this paper is that the Bayesian hierarchy must be a **dynamic hierarchy**, with dynamic temporary coalitions of inference units. The static mapping in the table above is ill-suited for this, in two main respects:

- Scaling of connectivity costs
- Binding of features to inference units

The dynamic hierarchy is determined by **spatial geometry**: two inference units can share a feature only if it might occur at the same spatial location for both of them. A new computational and linking component, the **aggregator**, is proposed to meet these requirements.

## 2. Knowledge Sources in the Brain

This section describes some background assumptions about knowledge sources in the brain, which underlie both the distributed and the aggregator models.

When using sense data to understand the positions and motions of objects in local space, many different types of constraint may be applied. Some of these are:

- edge detection
- motion detection
- stereopsis
- sound location
- locations of touch and movement sensations
- shape from shading
- shape from motion
- linking data from two or more sense modalities
- recognition of learned shapes or movements
- knowledge of hierarchical structures, such as bodies and body parts

The possible locations in space of objects and features are constrained both by raw sense data (vision, sound, touch) and by all the types of constraint listed above. It appears that animal brains can often make a near-Bayesian maximum likelihood estimate of the positions of features, in the light of all these constraints.

The different types of constraint appear to be applied by different regions of the brain, typically in the neo-cortex.

This leads to the idea that the capability to apply each type of constraint (or to learn it) resides in a limited region of the cortex. These modular regions will be referred to as **knowledge sources**.

There is a division of knowledge sources into primary and secondary knowledge sources:

1. **Primary Knowledge Sources** include incoming sense data, such as vision, sound and touch; and they could be defined to include early processing of sense data, such as edge detection.
2. **Secondary Knowledge Sources** include all other constraints such as shape from motion, shape from shading, and the recognition of learned spatial patterns such as faces or body plans, and use of hierarchical structure. They also include the linking of data from different senses - such as linking a sight with a touch sensation. A secondary knowledge source links together a cluster of features which may occur at any spatial displacement relative to the animal. It defines constraints on the relative positions of different features.

Some assumptions about secondary knowledge sources:

- There are several different types of secondary knowledge source, as described above.
- Each type of secondary knowledge source may involve one or more modalities of sense data.
- For each type of secondary knowledge source, a variable number of instances may be active at any moment. For instance, for a 'face recogniser' knowledge source, several instances may be active at one moment, recognising different faces. Several Shape from Motion (SFM) knowledge sources may be active at one time, for different parts of the visual field.
- One instance of a secondary knowledge source constrains a fairly small number of features. Some of these features may depend on other secondary knowledge sources, in a hierarchical manner. For instance, a face recogniser may use a nose feature, from a nose recogniser.
- Each secondary knowledge source constrains only the relative positions of the features it knows about. The face recogniser constrains the relative locations of nose, eye and mouth features.

One type of knowledge source is particularly important in motivating the considerations of this paper. This is **Shape from Motion** - (SFM) the ability to perceive rapidly from apparent movement that a set of features are part of the same rigid three-dimensional object. Shape from motion is important because:

- It is a vital knowledge source for small primates - both for moving about in trees, and for understanding the moving body parts of other animals
- At any moment, several parallel SFM calculations may be needed, for different parts of the visual field
- SFM requires rapid and precise computations of displacements in three dimensions
- In these computations, depth matters - depth is not a 'third class citizen' compared to the two dimensions of the visual field.
- Sets of features, which are near to one another in the visual field, need to be rapidly and dynamically assigned to different rigid bodies, in temporary groups of features.
- These rapid 'where' computations need to be dynamically linked to the 'what' computations of object recognition

Shape from motion therefore motivates two key ideas of this paper: (a) the need for precise 3-D geometric computations in the brain, and (b) the need for transient dynamic coalitions of knowledge sources.

Shape from Motion is discussed more fully in section 8 of the paper.

### 3. Computational Scaling Factors

A key objective of this paper is to discuss how the costs of spatial computations scale with various parameters of the mammalian brain (here called scaling factors) which can be estimated or measured. Having analysed the scaling properties of different models of computation, we can then compare how the costs of the different models scale as those factors vary.

I define here the main scaling factors involved, and make approximate estimates of their values for a small primate such as a macaque monkey. These are very approximate estimates, and could easily be refined; but even within their present large uncertainties, they lead to interesting comparisons between the models.

The scaling factors, their estimated values, and a brief rationale for the proposed values follow:

| Scaling Factor | Symbol | Typical value | Rationale |
|---|---|---|---|
| Number of features being tracked in any second | F | 20 | A small mammal needs to know the location and nature of 6 or more objects at any instant, in order to know where to move next. Assume that 3 features are needed per object. |
| Number of spatial dimensions | D | 3 | Some KS (such as shape from motion, shape from shading, object recognition) depend essentially on depth. Some species are not visually dominated. So D = 2 is not enough. |
| Number of secondary knowledge source types | T | 100 | Learned pattern recognisers for known shapes (bodies. limbs, plants, parts of all these) are counted as separate KS types. There are many such shapes, A hierarchy of 3 or 4 levels builds up the estimated number 100. |
| Number of knowledge source instances | K | 400 | There is space in cortex for at least this number of cortical modules. There are many instances of primary and early KS. Sometimes an animal needs several instances of one type of KS - e.g. several things of the same type may be present. |
| Number of knowledge source instances active at one moment | A | 30 | This number is similar to F, the number of distinct things whose identity and location needs to be tracked. Most pattern recognition KS are not active at any moment, because the things they recognise are not present. |
| Average number of features whose relative locations are constrained by one knowledge source | f | 5 | Hierarchical pattern recognition implies that the pattern to be recognised at each level of the hierarchy is not very complex. |
| Spatial resolution of relative positions of features in one KS | r | 1 in 5 | Hierarchical pattern recognition does not require very high resolution at any level of the hierarchy |
| Range of distances and scales over which one object needs to be recognised | s | 10 | Small mammals can recognise objects over this range of distances from themselves. |
| Average number of 3D constraints per feature being tracked | c | 8 | The total number of 3D constraints active at any moment is $Af = Fc$. |
| Number of re-estimations of feature locations per second | n | 10 | Small mammals require to update their model of local space as rapidly as this, in order to control their movements. |

One conclusion from this table is that there are many more instances of knowledge sources available in the cortex (400) than are in use at any moment (30). This applies particularly to secondary knowledge sources.

You may feel that some of these scaling factors should have different values from those shown above. If so, please use your own preferred values in the discussions which follow. There is possibly a useful research project in tracking some small mammal through the early stages of its life, to obtain better estimates of the scaling factors.

## 4. The Distributed Model

We next discuss the scaling and other properties of a fully distributed model of spatial cognition, such as many hierarchical Bayesian models of vision, or Rolls' (2015 and previous) VisNet model.

While these models differ in their detailed implementations, we can abstract some general shared features:

- There is a hierarchy of knowledge sources, typically with about 4 levels, applying constraints at different levels and of different types.
- All communication is done directly between knowledge sources, not through any shared central component
- Implicitly, the connections between knowledge sources (which may be reciprocal) are identified with cortico-cortical axons; there is a direct 'hard-wired' axonal connection from one knowledge source to another.
- A Bayesian 'level of belief' in the hypothesis of a specific knowledge source is encoded in the firing rates of its output axons
- Each knowledge source is sensitive to spatial information about the relative displacements of features coming from other knowledge sources
- Two-dimensional spatial information is encoded in the 2D structure of a sheet of neurons, or of the bundle of its efferent axons
- Any information about a third spatial dimension of 'depth' is encoded in some other way, often not specified

I next consider how distributed models address the requirements (a) - (e) described above

**(a) Scaling of Computational Costs**

From the scaling factors listed in the previous section:

- The number of active instances of knowledge sources is $A = 30$.
- The number of features being tracked at any moment is $F = 20$.
- The spatial location of any one of these features is jointly determined by $c = 8$ of the active knowledge sources.

In a Bayesian model for determining the location of one feature, each of the 8 knowledge sources estimates a most likely position for the feature, and estimates the shape of a likelihood function around this maximum. An additive combination of the negative log likelihoods leads to an estimate of the most likely position for the feature, in the light of the 8 knowledge sources.

It follows that the knowledge sources need to communicate and collaborate among themselves to determine the most likely position. If any one of them failed to communicate with the others, its estimate of the position would be ignored. In the fully distributed model, this requires up to 8*7 = 54 reciprocal communication paths between the KS.

In the distributed model, therefore, there need to be **temporary coalitions** of a few active knowledge sources, to determine the most likely position for each feature being tracked.

The need for temporary coalitions does not fit easily with the idea of hard-wired axon connections between knowledge sources. Recall the scaling assumption that there are 400 or more available knowledge sources. If the connections between them are permanent direct bundles of axons (each of which can be switched on or off to make temporary coalitions), each knowledge source needs to have outgoing connections to 400 others; any of these 400 connections might be needed at any moment for some temporary coalition.

So if the connections between knowledge sources were hard-wired direct neural connections, each knowledge source would need to have such connections to 400 other knowledge sources, but fewer than 30 of these connections would be in use at any time. Regardless of detailed numbers (which may be questioned) this appears to be less economical than a different architecture with some intermediate 'switching' component to lessen the number of required links (but which would contradict the assumption of a fully distributed model); and as we shall discuss below, the required bandwidth of the 400 connections per knowledge source might make the required number of cortico-cortical axons prohibitive.

**(b) Feature Binding**

A typical statement of the feature binding problem (Treisman 1998; Feldman 2013) is: how does any part of the brain know that two different features of the same object (such as redness and squareness) should be bound together to describe one object, and are not aspects of separate objects?

In terms of the fully distributed model, the question can be re-phrased: if a temporary coalition of 8 knowledge sources is needed to fix the location in space of one feature, how are those knowledge sources bound together to do that? How do they share information about that feature, without confusion between features?

This question can be split into two successive questions. In an example:

- What is the information that determines that knowledge sources 31, 67, and 204 need to be bound together to fix the position of feature 16? Where is that information held, and where is the required set of KS computed?
- Once it is established that knowledge sources 31, 67, and 204 need to be bound together, how is that realised in terms of neural connections?

Consider the first question from the viewpoint of knowledge source 31. It is constraining the location of feature 16 - but it has no information that knowledge sources 67 and 204 are constraining the same feature. So it cannot possibly compute the need to link together with 67 and 204 in a coalition - or to activate its hard-wired links to those KS. None of the distributed knowledge sources have the necessary information to know the need for a specific coalition of KS, let alone to implement it.

We might suggest a solution in which every feature was classified in a some 'alphabet' of feature types A, B, C.... Each KS would encode and broadcast the feature types it has detected, and would only be sensitive to incoming features of the same types. Thus, groups of KS interested in the same feature type B could form a coalition. However, this coalition would fail to distinguish between features of the same type coming from different spatial locations, as in the well-known 'Red triangles and blue squares' tests of feature binding. Similarly, binding by synchrony is a very weak form of binding, with insufficient information capacity to define the KS coalitions required - capable of handling only a very small alphabet of feature types (Shadlen & Movshon 1999).

There seems to be a logical contradiction between a fully distributed model, and the need to form temporary coalitions of KS. Some component other than the knowledge sources is needed (a) to hold the information that defines which KS need to collaborate, (b) to compute from that information which KS need to collaborate, and (c) to make them collaborate.

**(c) The need for precise spatial computations**

A widely used example of hierarchical Bayesian Knowledge sources is the example of noses and faces. A 'nose recogniser' KS infers the likely existence of a nose, and passes this up the hierarchy to a 'face recogniser' KS. If there is high confidence in the existence of a face, this feeds back in a top-down manner, to increase the level of confidence in the nose.

However, this mutual reinforcement of Bayesian likelihoods should only happen if the nose occurs at the correct place in the face. A nose in the wrong place - as in a Picasso face - should not produce reinforcement.

While I have cited the face/nose example for its familiarity, it is not the best possible example, because the importance of face recognition has led to the evolution of specialised face recognition regions in the primate cortex - where a nose is always a part of a face. A better example would be the recognition of classes of item with no evolutionary history, such as cutlery - where a fork might have a three-level hierarchy of parts as in: fork - fork head - prong. The large number of such hierarchical structures, and lack of pre-determined whole/part relationships, makes forks a better example than faces.

How is this requirement for correct displacements between the whole object and its parts computed, in the fully distributed model?

Individual knowledge sources are mainly concerned with the relative positions of their constituent features. Thus the fork KS is mainly concerned with the relative positions of handle, head, and prongs; and the handle KS is concerned with relative locations of parts of the handle.

If the handle KS knew only about relative displacements in a handle, and the fork KS knew only about relative displacements within a fork, there would be no possible way to convey information from the handle KS to the fork KS about where the handle is in the fork; between the two KS, the required information simply does not exist.

Therefore we are forced to assume that the handle KS also knows about (or computes) the absolute position of the handle in some frame of reference; and the fork KS knows about or computes the absolute location of the fork in the **same** frame of reference. Only then would the required communication between the two KS, about the position of the handle in the fork, be possible at all.

In order for this to happen, there needs to be some precise computation of three-dimensional spatial displacements. This is shown in the diagram below.

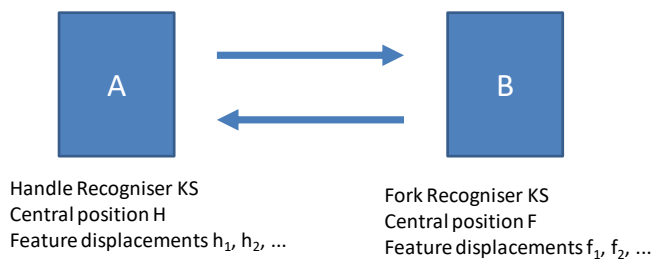

Handle Recogniser KS
Central position H
Feature displacements $h_1, h_2, \ldots$

Fork Recogniser KS
Central position F
Feature displacements $f_1, f_2, \ldots$

The 'Handle' KS computes the absolute location of the handle - denoted by **H** - and the relative displacements $\mathbf{h}_i$ of features in the handle. The Fork KS computes the absolute location of the fork - denoted by **F** - and the relative displacements $\mathbf{f}_i$ of features in the fork. One of these features is the handle.

The 'Fork' KS needs to know the relative displacement of some part of the handle - say the end of the handle, $\mathbf{h}_0$ - in the frame of reference it uses for its own local relative displacements. In other words, it needs to know the small vector

$$\mathbf{h'}_0 = \mathbf{h}_0 + \mathbf{H} - \mathbf{F}$$

The only way to compute this vector in a distributed manner - with each KS using only the information that is available to it - is as follows:

1. The handle KS computes and transmits $\mathbf{m}_0 = \mathbf{h}_0 + \mathbf{H}$
2. The fork KS computes $\mathbf{h'}_0 = \mathbf{m}_0 - \mathbf{F}$

Therefore, the position information exchanged between the two KS cannot be just about the small relative displacements which they use internally; it must be about absolute displacements in some shared frame of reference. This implies that the calculations of displacements (1) and (2) above - and the communication between the two KS - must have high spatial precision.

In essence, **H** and **F** can be large spatial vectors - such as the displacement of the fork from the viewer - whereas $\mathbf{h}_0$ and $\mathbf{h'}_0$ are smaller vectors - displacements within a fork or a handle. But they are computed as the difference of two large vectors.

For instance, if recognition of a fork requires a precision of 1 part in 5 in the local displacements, but forks can be recognised at an absolute displacement of up to 10 fork lengths from the viewer, then all the spatial computations and communications between KS all need to be done with spatial precision of 1 part in 50.

This is a very demanding requirement for small, localised KS distributed across the cortex.

It might be thought that, at least in two dimensions, the necessary spatial precision could be got by using small sheets of neurons, and encoding position as the location of a high firing rate within the sheet. Even this would be demanding, for small localised knowledge sources. But there are arguments that two dimensions alone are not sufficient:

- Many important knowledge sources - such as shape from motion and shape from shading - are dependent on precise storage of depth information
- For animals less visually dominated than primates, the two dimensions of vision are not pre-eminent
- There is often no preferred frame of reference which will work across diverse KS including vision and touch

As well as the difficulty of representing and computing 3D spatial displacements precisely within the many small modules in cortex, there is also a high cost of communicating this precise information between knowledge sources. To give high spatial precision in sub-second timescales, a high bandwidth between any pair of knowledge sources is required. This would seem to require prohibitively large numbers of cortico-cortical axons - most of which would not be used for most of the time.

We have shown that, even to allow knowledge sources to communicate small relative displacements between themselves, they need to compute and communicate larger absolute displacements, in some frame of reference which is common to all of them. The need for this common frame of reference is another indicator of the need for a separate component of the architecture as well as the knowledge sources. It is not possible for one knowledge source on its own to compute what the shared frame of reference should be. It is also hard (although not impossible) to see how the shared frame of reference could be rapidly and precisely computed by some distributed cooperation of the knowledge sources. This seems to point to the need for a separate component in the architecture, to compute the shared frame of reference and promulgate it to all knowledge sources. Without this component, they cannot share spatial information.

The need for precise spatial displacements shows that there is a need not only for **switching** between knowledge sources (to make a temporary coalition of KS 35, 46, and 101) but also for precise **steering** of feature locations between them (to align feature 17 at the appropriate position in all KS in the coalition). The need for both switching and spatial steering places yet more strain on the idea of permanent direct neural connections between KS; fine steering would require yet more fixed neural paths to select from.

Olshausen, Anderson and VanEssen (1993, 1995) have considered a switching architecture, with switchable neural pathways between cortical knowledge sources. Out of many available pathways, one pathway is dynamically selected at any time, depending on the steering requirement. Clearly, the higher the spatial precision required in the signal steering, the larger the number of switchable paths there must be; and high spatial precision in three dimensions would seem to require large numbers of paths to be available between any two knowledge sources.

This can be seen in a simple example where switchable pathways are required for two-dimensional tasks (e.g in vision) with a spatial precision P – that is, where 2-D relative positions are defined with a precision of one part in P. Primates might require P in the range 5 – 40 for hierarchical visual recognition. What is the cost of switchable pathways between two cortical 'patches' A and B, at neighbouring levels of the hierarchy, in a hierarchical cortical model like that of Olshausen et al.? Each patch A and B requires $P^2$ cortical columns to store its spatial information. For the paths between A and B to be steerable for any 2-D spatial displacement, with precision P, there need to be $P^2$ pathways, each containing $P^2$ axons. This makes a total of $P^4$ axons, even in the two-dimensional case.

For three-dimensional tasks like shape from motion, the number of axons would be larger. Similarly, if absolute positions were required in some animal-centred frame of reference, the required precision P would be larger.

$P^4$ growth is very rapid, and soon becomes prohibitive. A growth in axons and brain energy costs like $P^4$ would soon be untenable, for animals requiring greater precision of spatial cognition, if there is any alternative. The aggregator model described below is such an alternative.

**(d) Fast Learning of Invariant Spatial Patterns**

Most of the pattern recognition knowledge sources in the hierarchy need to learn the patterns they recognise, from the smallest possible number of training examples.

The fastest possible speed of learning is constrained by a Bayesian limit on learning, which shows that the smallest number of training examples required to learn some rule grows as the log of the number of possible rules that might be learned.

Consider for instance a grid of 5*5 cells, each of which has a binary input - on or off. The number of possible input patterns on this grid is $2^{25}$. There is a set P of possible patterns, whose size is $2^{25}$. To learn a rule is not to learn a single pattern, but to learn a subset of the patterns which has some significance. For instance, a rule might be 'Any pattern within the subset S of patterns indicates a type of food'. The learning problem is then to infer the set S from training examples. The number of subsets S of the set of P patterns is $2^{**}(2^{25})$; so without further constraints, the number of training examples needed to learn the one correct S is of the order of $2^{25}$, or 30 million. Further constraints are necessary.

This example illustrates that for fast learning in the course of a single lifetime, two things are needed:

1. A small initial space of patterns which might be learned
2. Powerful Bayesian priors about what sets of patterns in this space might be useful.

To illustrate the requirement for a small initial space of patterns - if faces are projected in an aligned fashion on a 5*5 grid, learning can be much faster than if they are projected at random places on a 10*10 grid. Fast learning of spatial patterns requires accurate alignment of the pattern to be learned on some input grid of the learning module.

This reinforces the need for accurate spatial steering of patterns to be learned from incoming sense data onto the learning knowledge sources. It gives another reason why precise computation of the steering displacements - to align spatial stimuli precisely on learning modules - is an essential part of spatial cognition.

**(e) Exploiting Object Constancy**

Animal brains have evolved to make good use of the Bayesian prior probabilities which prevail in their habitats. This can happen only if those priors have been true for long enough to have shaped the evolution of the animals' brains.

There is one Bayesian prior probability which has been true for all evolutionary time, and which can therefore be expected to have had very important consequences for animal brains. This is:

> *Most of the time, most things do not move.*

The importance of this constraint can be seen by imagining an animal whose brain could not make use of it. There would be two important consequences:

- Rather than assume that: 'the rock was over there 5 seconds ago; so in the absence of other sense data, I will assume it is still there' the animal would need to be constantly checking and re-checking that the rock is still there. This would lead to a greater workload on its sense organs and spatial brain, putting it at a big disadvantage.
- An important principle of attention is: 'if something moves, it needs attention'. What moves might be a predator, or food. If the animal's brain had no way of determining and filtering out all things which are not moving (in spite of the animal's own motion, visual saccades and so on) it would have no way to steer its attention towards those things that are really moving, and which need attention.

So a brain which cannot easily determine that: 'that thing has not moved' is at a severe disadvantage. Empirically, we know that animal brains, including our own, are good at distinguishing real movement of objects from changes in sense data driven by one's own movement.

In a fully distributed model, where knowledge sources deal in relative displacements, it is hard to see where in the brain this happens. This is another indicator of the need for a separate component to the brain architecture, as well as the distributed knowledge sources.

We saw under topic (c) that in order for knowledge sources to communicate small relative spatial displacements between themselves, they need to compute and communicate absolute displacements in some common frame of reference. There needs to be some separate component to the architecture, to compute and promulgate the common frame of reference.

If this common frame of reference was a truly static frame - one in which non-moving objects do not appear to move - then it would be easy for each knowledge source to detect whether any of its set of input features was actually moving. But this can only happen if there is a separate component in the architecture, which computes the static frame of reference.

## 5. The Aggregator Model

We next ask whether there is an alternative to the fully distributed architecture, which allows the constraints from separate knowledge sources to be combined in a near-Bayesian manner, but which is better placed to meet the requirements (a) - (e).

In the aggregator model, there are secondary knowledge sources, each of which applies constraints such as shape from shading to a small set of related features. There is also an **aggregator**, which stores the best estimate of the position of each feature.

The model is shown in the diagram:

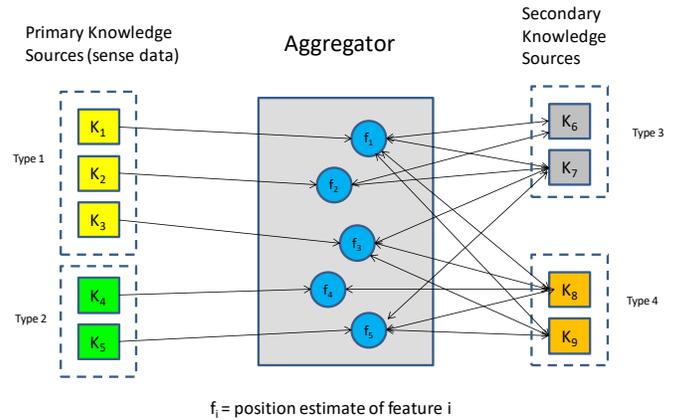

$f_i$ = position estimate of feature i

In this diagram, the primary knowledge sources $K_1 .. K_5$ place constraints on the positions of features $f_1 ... f_5$. Each secondary knowledge source $K_6 .. K_9$ places constraints on the relative positions of some small subset of the features. The dashed boxes group different types of primary knowledge sources (e.g. different modalities of sense data) or types of secondary knowledge sources (such as shape from shading). The arrows show communication paths. The aggregator stores the absolute positions in three dimensions of all features, in a frame of reference to be discussed below.

Tracking the position of each feature can be done by an iterative two-phase **aggregator cycle**, with information flowing along the arrows in the diagram. The two phases are:

1. Each knowledge source, primary or secondary, independently applies its own constraints to the positions of a small set of features - and passes the resulting positions in to the aggregator (the KS phase).

2. For each tracked feature, the aggregator combines the position estimates from each knowledge source - and passes the resulting positions back out to the secondary knowledge sources (the aggregator phase).

The knowledge sources do not communicate position information directly with one another, but communicate positions only through the aggregator. They may communicate other information directly.

The aggregator cycle is illustrated in the following diagram:

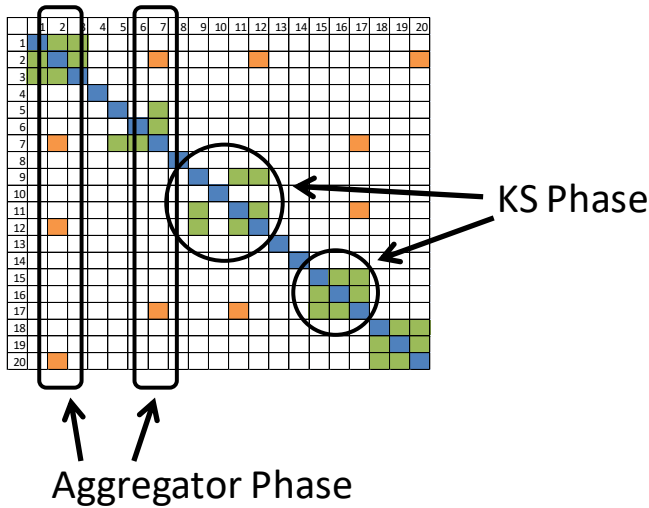

The symmetric 20*20 matrix shows all the constraints from all active knowledge sources when tracking 20 features. Rows and columns are features, and a coloured cell indicates a constraint on the positions of a pair of features. Blue diagonal cells arise from primary knowledge sources. Green near-diagonal cells are constraints on nearby features, from single knowledge sources. Orange cells, far from the diagonal, are constraints which arise when two or more distinct knowledge sources constrain the same feature - for instance, when touching and seeing the same object. These are constraints like: 'the pencil I see is at the same place as the pencil I am touching'.

If the likelihoods from all knowledge sources were Gaussian ($\sim \exp(-x^2)$) near their maximum, then each constraint cell denotes a quadratic log likelihood around some most likely relative displacement. The total log likelihood is a quadratic form. Finding the optimal Bayesian configuration of F features would require solving 3F linear equations - an expensive process, and it is not clear how it would be done for large F.

The aggregator cycle splits this into a number of smaller parallel computations. In the KS phase of the cycle, each KS independently finds the best local fit to its own constraints (circles above). In the aggregator phase (vertical rounded squares), for each feature the aggregator combines the estimates coming from each KS into an overall best estimate. This process converges rapidly to a near-Bayesian fit.

This cycle allows a high degree of parallelism:

- Each secondary knowledge source can compute its own constraints in parallel with all the others
- In the aggregator, the combined position estimate for any feature can be computed in parallel with that for any other feature

For each knowledge source to make a Bayesian estimate of the positions of the features it relates, it needs to have prior estimates of the precision (or uncertainty) of the position of each feature before it computes. These estimates are passed from the aggregator to the knowledge source.

Similarly, in order for the aggregator to make a Bayesian estimate of the position of each feature, the precisions in the position estimates of the feature arising from each knowledge source must be passed to the aggregator.

Therefore, estimates of the precisions in positions (in three dimensions, the precision tensors) of features are passed round the iterative cycle between the aggregator and the knowledge sources - just as the positions themselves are passed round the cycle.

The cycling of precisions must be managed carefully, to avoid a cyclic escalation of spuriously increasing precisions. I have done this in a computer model of the aggregator cycle. This model gives estimates of the positions of objects, and of their uncertainties, which are approximately the same as a full Bayesian maximum likelihood estimate, although they are not exactly the same.

The aggregator cycle gives an economical and well-defined iteration to track the most likely positions of a set of features, in the light of Bayesian constraints on those positions from a large set of (~30) independent active knowledge sources. Without the aggregator cycle, it is hard to see how the whole set of knowledge sources would collaborate to find a near-Bayesian solution. This might be done by a 'grand coalition' of the knowledge sources, larger than the coalitions needed in the aggregator model; but it is not clear how rapidly the solution would converge.

I next assess the aggregator model against the same criteria (a) - (e) with which I assessed the distributed model.

**(a) Scaling of Computational Costs**

The most obvious difference between the two models is this: while the distributed model requires exchange of spatial information directly between knowledge sources - each knowledge source being able to pass information directly to every other one, or to a large subset of them - in the aggregator model, each knowledge source need only exchange information about positions of features with the aggregator. This leads to a dramatic reduction in the required number of communication paths, effectively from $K^2$ to K (where we have estimated K = 400). This translates directly to a saving in net axon length, and thus to a major saving in brain energy costs.

The comparison of costs of computation is not so clear-cut. In both models there is some kind of iterative relaxation towards a best-fit model of spatial positions, against a backdrop of changing spatial positions and changing sense data. A comparison of levels of performance of these two relaxation processes, in terms of precision versus required number of computation operations, is difficult to do without defining more precisely the fully distributed iterative model.

Both models require the precise computation of spatial displacements. The level of precision of the computation, and the number of computations required in every second, seem to be the same in both models. In terms of the scaling factors described in section 3, the number of spatial displacement computations per second is of the order of ncF = 10*8*20 = 1600.

It is clear, however, that the aggregator model supports a high degree of parallelism in the computation - whereas in the fully distributed model, the allowed degree of parallelism is less clear.

**(b) Feature Binding**

When considering the fully distributed model, we saw that feature binding required the solution to two problems:

- For each feature whose location is being tracked, deciding which small set of knowledge sources need to cooperate in a temporary coalition to fix the position of that feature
- Once the temporary coalition has been defined, communicating between the knowledge sources in the coalition, to determine the location of the feature

In the distributed model, both of these problems are difficult. The aggregator model suggests an answer to both problems, through the **spatial binding of features**.

The aggregator is a shared store for the estimates of the positions of features coming from all knowledge sources, which then pools the estimates of the position of any one feature. To do this, the aggregator must store the position estimates of features effectively indexed by their spatial position in some shared frame of reference - so that spatially close feature estimates are stored in a way that is close in the aggregator, and the aggregator can recognise this proximity.

Then the set of knowledge sources which need to be bound together, to give a joint estimate of the position of some feature, is just the set that give spatially close (overlapping) estimates of its position. The aggregator can detect if two position estimates are spatially close, because it spatially indexes features by their location. This is the way the aggregator computes the temporary coalition of knowledge sources needed to locate some feature. Then, binding the estimates together to make a joint estimate of the position of the feature is the other function of the aggregator. Collaboration of the temporary coalition is done within the aggregator.

This theory of the spatial binding of features has much in common with the Feature Integration theory of Treisman & Gelade [1980] - in which features are indexed by their locations, in a feature map. The aggregator is the map.

This seems at first to be putting a heavy workload on the aggregator - requiring it in effect to solve the binding problem. There are possible neural computational mechanisms to do it, which will be described in a future paper, and are sketched in section 7 The aggregator separates out the binding problem from the other problems being solved by local knowledge sources in cortex, and makes it easier to solve in a dedicated structure.

**(c) Computing Precise Spatial Displacements**

In the distributed model, in order for any two knowledge sources to exchange information about the positions of features, they need to exchange not just small relative displacements of features, but to exchange more precise absolute positions in some shared frame of reference.

The same is true in the aggregator model, and the model suggests how the requirement may be met:

- In the KS phase of the aggregator cycle, each knowledge source sends to the aggregator the estimated relative positions of its features.
- The aggregator needs to store absolute positions in some frame of reference. It knows the absolute position of each knowledge source in that frame of reference, and uses that information to convert the relative displacements of features, as sent by the knowledge source, into absolute position estimates in its single frame of reference. This is a 3-D vector addition of spatial displacements.
- The aggregator then recognises and combines the nearby estimates of the position of any feature, to get a pooled estimate of the absolute position of that feature.
- When sending the pooled position estimates out to the knowledge sources, the aggregator uses information about the absolute centre of each knowledge source, to convert absolute feature positions to relative displacements in the knowledge source.

Comparing this proposed aggregator solution with the distributed solution:

1. The aggregator model does not remove the need to compute precise three-dimensional displacements; but it allows this computation to be done inside the aggregator, which can be a large special-purpose structure evolved to do spatial computations.
2. The only displacements that need to be communicated over long distances in the brain (between KS and the aggregator) are not absolute positions, but are smaller relative displacements within a knowledge source. These small relative displacements do not need to be communicated with the same high precision and bandwidth as absolute positions would require.

Factor (2) leads to a further saving in brain energy costs of the aggregator model, compared to a distributed model.

**(d) Invariant Pattern Learning**

In either distributed or aggregator brain architecture, to support fast learning of spatial patterns, incoming sense data needs to be precisely steered onto the input neurons of learning knowledge sources. From the discussion under (c), it should be clear that the aggregator performs just this function, and is a specialised structure for doing it.

**(e) Exploiting object constancy**

In order to recognise that some object is not moving, the brain needs to have some storage structure in which the lack of movement is represented. Cortical knowledge sources, being concerned with relative displacements, are not this structure. This was a problem for the fully distributed architecture, and led to the need for some central component to define a stationary frame of reference.

The aggregator needs to store absolute positions of all tracked features in some single shared frame of reference. If it is designed so that this single frame of reference is for most of the time static (so that stationary objects do not appear to move), then the aggregator can serve as the place in the brain where the truly static is recognised as static; and it can be the short-term memory for those locations, to avoid having to re-compute them from sense data, and to detect real movement which requires attention.

**Summary of requirements**

Summarising across the requirements (a) - (e): in each case, the aggregator architecture appears to have significant advantages over the fully distributed architecture. These are not just advantages of scaling and cost; we have seen that the distributed architecture logically needs some component other than the knowledge sources, in order to define absolute positions of objects.

## 6. Where is the Aggregator in the Brain?

If the hypothesis of an aggregator in the mammalian brain is correct, then there is a simple argument that the aggregator must be physically close to the centre of the brain.

This is because the aggregator needs connections from primary knowledge sources, and reciprocal connections to all secondary knowledge sources, which are distributed around the approximately spherical cortex. If the aggregator is near the centre of the sphere, then the net axon length of all these connections is minimised. So a near-central position will minimise the axonal energy consumption of the brain; brains will evolve towards that configuration.

The requirement for a central physical position implies that the aggregator is not any region of the cortex.

The most prominent structure near the centre of the brain is the thalamus. There are other reasons to suppose that the aggregator is the thalamus:

- The thalamus acts as a gateway to cortex for sense data of nearly every modality. These modalities are used to fix the locations of objects near the animal - which is the core function of the aggregator.
- Notably, the thalamus is not the gateway to cortex for olfactory data. This makes sense, because smells are of very little use in fixing the locations of things - unlike other sense data, smells can come from anywhere. It would make little sense to pass smells through the aggregator.
- The thalamus is richly connected to cortex, and has all the required connections to act as an aggregator. It has incoming connections from sense data (primary knowledge sources) and has reciprocal connections to many regions of cortex, as is required to connect to secondary knowledge sources.
- Waking thalamo-cortical rhythms include high frequencies around 40 Hz, or alpha frequencies around 10 Hz, either of which is a suitable frequency for the aggregator cycle, to track and re-compute positions of objects many times per second
- The thalamus has a prominent and fairly conserved form across many species. This is consistent with an important role of determining the positions of things many times per second
- Each half of the thalamus has an approximately spherical shape (it is extended in all three dimensions) which may be suitable to hold a distributed high-precision representation of spatial positions (see section 7)

With this identification of the thalamus as the aggregator, the aggregator model is a special case of Mumford's [1991] proposal that the thalamus acts as a 'blackboard' (Erman et al [1980]) collating information about features from different cortical knowledge sources. In this case, the thalamus specifically collates information about the locations of features.

## 7. Possible Neural Models of the Aggregator

The aggregator model avoids many of the difficulties of a fully distributed model of spatial cognition, by proposing an aggregator which plays a key role in spatial cognition and feature binding. It might be thought that this is just a way to collect all the hard problems in one place, rather than to solve them.

If there appeared to be no possible neural implementation of the functions proposed for the aggregator, then this criticism would carry some weight. If, however, one could describe in outline how an aggregator might work, then the criticism would not carry weight - and there could be a useful research agenda in turning the outline into a working model.

A detailed consideration of how the thalamus might function as an aggregator is the subject of another paper, in preparation. This section sketches at a high level two possible ways in which an aggregator of location information could work.

Some key requirements for a position aggregator are the following:

1. Ability to store the absolute locations of a number of features currently being tracked (of the order of 30 features, with multiple estimates for each feature from different knowledge sources)
2. Ability to store these locations with high spatial precision (e.g. better than 1 part in 50) in three dimensions
3. Ability to recognise that two estimates of the position of feature are spatially close to one another, and so need to be bound together (i.e. treated as estimates of the position of one feature)
4. Ability to compute spatial displacements (vector differences between stored positions) rapidly and with high precision.
5. Ability to combine two or more estimates of the position of a feature into a 'pooled' estimate

The core requirements (1) and (2) are to represent spatial positions with high resolution in all three dimensions. This could be done by a Fourier transform-like representation along the following lines:

Consider an 'input' neuron whose output synapses are spread spherically around some point in the brain denoted by **c** = **0** (using bold characters for three-dimensional vectors), where the density of synapses at any point **r** is given by:

$$\rho(\mathbf{r}) = \exp(-\alpha \mathbf{r}^2) * [1 + \cos(\mathbf{k}.\mathbf{r})] \qquad (1)$$

This is a wave-like striated pattern, with planes of maximum density of synapses (or of minimum density, zero) perpendicular to the vector **k**. The distance between two planes of maximum synapse density (the wavelength) is $(2\pi/|\mathbf{k}|)$.

A high firing rate of this neuron will represent an object at position

$$\mathbf{x} = \beta \mathbf{k} \qquad (2)$$

where $\beta$ is some constant.

The neuron represents an object at the position $\beta \mathbf{k}$ in the following sense: suppose there are 'output' neurons with the synapses from the 'input' neurons as their input synapses. One of these output neurons has a total density of excitatory synapses (from all the input neurons):

$$\rho_e(\mathbf{r}) = \exp(-\alpha \mathbf{r}^2) * [1 + \cos(\mathbf{k'}.\mathbf{r})] \qquad (3a)$$

and a total density of inhibitory synapses:

$$\rho_i(\mathbf{r}) = \exp(-\alpha \mathbf{r}^2) * [1 - \cos(\mathbf{k'}.\mathbf{r})] \qquad (3b)$$

Weighting the excitatory synapses +1, and the inhibitory synapses -1, gives a net density of synapse excitation

$$\rho_i(\mathbf{r}) = 2 \exp(-\alpha \mathbf{r}^2) * \cos(\mathbf{k'}.\mathbf{r}) \qquad (4)$$

The density of synapses from one input neuron to one output neuron is the product of (1) and (4). Setting aside the constant term, this product is of a form

$$\exp(-2\alpha \mathbf{r}^2) * \cos(\mathbf{k}.\mathbf{r}) * \cos(\mathbf{k'}.\mathbf{r}) \qquad (5)$$

which, when integrated over all **r** (all volume in the brain) has the form near **k** = **k'**

$$C \exp(-(\mathbf{k} - \mathbf{k'})^2 / \alpha^2) \qquad (6)$$

This means that signal transmission from an input neuron to an output neuron is strong only near **k'** = **k**; transmission is selective in the represented location.

Even if input neurons for several different locations $\mathbf{k}_i$ are active at the same time - representing features at all these locations - one output neuron will only be sensitive to features whose locations are close to its own represented location **k'**. Sets of input neurons can represent features at many locations at once.

This on its own would be of little use - acting just like a set of direct connections from the input representation of a position $\beta \mathbf{k}$ to an output representation of the same position. But now assume there are also 'steering' neurons whose output synapse density varies as

$$\rho(\mathbf{r}) = \exp(-\alpha \mathbf{r}^2) * [1 + \cos(\mathbf{M}.\mathbf{r})] \qquad (7)$$

where the vector **M** is called the steering displacement. Assume that the dendrites of any output neuron make a 'sigma-pi' summation of the products of the inputs from the input neurons (1) and the steering neurons (7). Then the summed strength of the signal to an output neuron is of the form

$$C \exp -(\mathbf{M} + \mathbf{k} - \mathbf{k'})^2/\alpha^2 \tag{8}$$

This function is large only in the region near $\mathbf{k'} = \mathbf{M} + \mathbf{k}$. So the output neuron will only fire strongly in the region close to

$$\beta\mathbf{k'} = \beta\mathbf{M} + \beta\mathbf{k} \tag{9}$$

The sigma-pi computation in the dendrites of the output neuron effectively computes the vector sum of two represented positions $\beta\mathbf{M}$ and $\beta\mathbf{k}$; that neuron will only fire strongly if its own represented position $\beta\mathbf{k'}$ is close to the vector sum $\beta\mathbf{M} + \beta\mathbf{k}$. This makes a precise and distortion-free computation of a spatial displacement - a vector difference.

The exponential expressions such as (8) for the sensitivity factors, which exhibit the spatial selectivity, are the results of integrating the synapse densities over the volume of the brain, as if there were infinite numbers of synapses to be summed over. In practice there is only a finite number N of synapses connecting any input neuron to any output neuron - so the actual strength of connectivity of neurons is not exactly given by this expression. The actual strength is like a Monte Carlo approximation to the integral, calculated with N random integration points, where N is the number of connecting synapses. Even for N = 100, the random errors in the Monte Carlo sum are not expected to be large (proportional to $N^{-0.5}$) - allowing any one output neuron, with for instance $10^4$ synapses, to connect to a moderate number of active input neurons, representing features at different locations.

This shows that if neurons in some extended three-dimensional region have synapse densities which vary as $\cos(\mathbf{k.r})$ within the region, as in (1) - (4), with a range of three-dimensional wave vectors **k**, they can represent features at positions $\beta\mathbf{k}$, and can compute spatial displacements between features (differences of vector positions).

The represented positions will have high spatial precision, provided the smallest dimension R of the region is large compared to the smallest possible wavelength $\lambda$ of the synapse density variation; spatial precision is approximately one part in $(R/\lambda)$.

The calculation of spatial displacements is precise and free of geometric distortions; and it can be done rapidly, within one neuron firing cycle. It does not require many firing cycles to accumulate high spatial precision, as it would do if components of positions (such as depth) were represented by neural firing rates.

Suppose some cortical knowledge source is served by output neurons tuned to a range of small relative displacements **d**. This knowledge source will not receive inputs from other spatial positions **D**, unless the steering displacement **M** is such that $\mathbf{D} = \mathbf{d} + \mathbf{M}$. So any two features $f_1$ and $f_2$ cannot be sent simultaneously to that knowledge source, unless their positions $\mathbf{d}_1$ and $\mathbf{d}_2$ are close to one another. Different features are bound by proximity of their represented spatial positions.

This form of spatial feature binding is highly selective - features can be bound into many distinct sets of features, because there are many regions of local space represented in the aggregator. Spatial binding also aligns well with the basic requirement for feature binding. We require features to be bound together if and only if they arise from the same region of real space - that is, if they are part of the same object or the same part of an object. So with this spatial binding, the computation of 'which features should be bound to which other features?' is in some sense made automatically, from their represented locations.

The Fourier representation can be used to encode not only an estimated position of some feature, but also the range of uncertainty of that estimate. The factor $\exp(-\alpha\mathbf{r}^2)$ in the equations above is a Gaussian falloff with increasing $|\mathbf{r}|$; if this is a rapid falloff (large $\alpha$) then the wavelength of the excitation has large uncertainty; while for small $\alpha$ the wave vectors **k** are well defined. So varying $\alpha$ varies the precision of the wave vectors which represent positions. The factor $\alpha$ is easily generalised to a tensor in **r**, which can express different degrees of uncertainty in different directions.

Then the Fourier representation can also be used to combine several different estimates of the position of one feature, coming from different knowledge sources, with their Bayesian weights depending on their degrees of uncertainty. This is because a Bayesian estimate is got by multiplying likelihoods, or equivalently by adding negative log likelihoods. If, near the best estimate position from each knowledge source, the signal strength from that knowledge source at each position is proportional to negative log likelihood it gives for that position, the addition of the signal strengths from different knowledge sources - which is done naturally by adding the firing rates of their neurons - produces an overall maximum at the position of the Bayesian best estimate. Adding firing rates in the Fourier representation of positions makes a Bayesian combination of the estimates, as long as the ranges of uncertainty are overlapping.

I have described in outline how if, in some extended three-dimensional region of the brain, there is a Fourier-like neural representation of the positions of features, it can meet the following requirements for the aggregator function:

- Precise storage and transmission of the spatial positions of many features
- Rapid and precise computation of spatial displacements
- Fine-grained spatial steering of input sense data from some small region of space to specific knowledge sources, by control of steering signals
- Spatial binding of features
- Bayesian combination of position estimates for a feature from several knowledge sources.

These can be the basis of a neural computational model of the aggregator function in the thalamus. Detailed definition of the model, and detailed calculations of the performance of the model, remain to be done.

The outline model sketched above is merely an 'existence proof' that precise storage of locations and precise calculation of spatial displacements could take place in the thalamus. The actual mechanisms used in the thalamus are expected to be more complex, using resources such as thalamic interneurons, thalamic compound synapses such as glomeruli (Sherman & Guillery 2006) and other complex dendritic computations (London & Hausser, 2005; Mel 2006 ). Many thalamic nuclei are spatially extended in three dimensions, so all of these could be the location of Fourier-like computations.

It is unlikely that a Fourier transform-like neural realisation of the aggregator functions could be done in any region of cortex, because cortex is only 2.5 mm thick; so it could not support a wave-like ($\sim\cos(\mathbf{k.x})$) distribution of synapses, with many wavelengths (as is required for high spatial precision) in the third dimension.

There is a second, more unorthodox suggestion for the representation of spatial positions of features in the aggregator. This is the proposal that feature positions are represented as the Fourier transform of a physical field in some region of the brain, and that neurons couple to that physical field (as they can couple, for instance, to light). A physical field suffusing the whole thalamus could be more economical than collections of synapses in making many-many connections between thalamic neurons; and it might reduce the level of random noise, compared to synaptic connections. The field could still support the precise calculation of spatial displacements, and the Bayesian summation of log likelihoods from different knowledge sources.

No physical field carrying spatial information has yet been found in the brain. This might be because the field is an exotic kind of excitation, or has very low intensity, or because we have not yet looked for it in the right way. Some features of thalamic anatomy support the suggestion that the thalamus is the site of a physical field (Worden 2010, 2014).

## 8. Shape from Motion

This section discusses a type of knowledge source which has been important in developing the ideas of this paper, and which lends support to them.

If a set of randomly-placed dots is shown on a computer screen, then when the dots are stationary, they are perceived just as random dots on a flat screen. But if their positions on the screen have been computed as the 2-D projections of the positions of points on a rigid three-dimensional body, then as soon as that body is rotated, the dots rapidly 'click' into a perception of the rigid body moving, with a vivid sense of depth. This happens regardless of the shape of the body, and even if the whole body subtends only a small angle in the visual field.

This simple experiment shows that the Shape from Motion knowledge source:

- Is independent of other knowledge sources, such as shape from shading or stereopsis - working well even if only given a few points of light on a screen
- Is robust and fast
- Does not depend on learned shapes - as it works for a large set of shapes, too large to be learned
- Is a pure geometric computation
- Requires sophisticated and fast spatial computation in three dimensions
- Requires high precision in the internal representation of positions, in all three dimensions.

It would seem hard to get the rapid and robust computation of 3-D shape from a representation of space by using only a 'two dimensions plus depth' in the visual cortex, with depth encoded in some way by neural firing rates. It would seem to take too long for the neural firing rates to define depths with sufficient precision.

The shape from motion knowledge source is biologically important not only in perceiving the shapes of things that are actually moving - but just as important, perceiving the shape of stationary objects (i.e. the majority of objects) as revealed by the animal's own movement. Identifying stationary objects is necessary for two reasons: (a) planning one's own movements around them, and (b) knowing by exception what is really moving, and so what needs special attention.

For instance, it seems likely that a small monkey moving fast in a tree, deciding how to move next, relies more on shape from motion than it relies on other depth cues such as stereopsis or Shape from Shading - particularly at the edges of its visual field. It seems likely that Shape from Motion needs to be capable of operating in parallel over many parts of the visual field, not just near the fovea; and that it often acts in collaboration with other knowledge sources.

The shape from motion capability on its own illustrates the need for temporary coalitions of knowledge sources. A set of features which are assessed to be part of a rigid 3-D body is in itself a temporary coalition of the knowledge sources constraining those features. When a 3-D shape from motion is used to recognise and classify an object, a further coalition is needed - between the shape from motion knowledge source, and a learned object recogniser knowledge source.

There are probably many instances of the shape from motion knowledge source active at one moment, and many distinct object recognisers (Leibo, Liao, Anselmi, and Poggio 2015). To allow them to collaborate directly in temporary coalitions would require many-to-many connections between shape from motion KS and object recogniser KS - with each connection conveying precise position information. Dynamic connections via some separate routing element are a more efficient way to do this, and one may ask whether static cortico-cortical connections have the required bandwidth to do it.

In a simple model of the cortex, one might expect shape from motion to occur mainly in the dorsal 'where' stream, and object recognition to occur mainly in the ventral 'what' stream; so that direct many-many connections between them would require high bandwidth links between these two streams. fMRI studies of shape from motion (Murray, Olshausen and Woods, 2003; Paradis et al, 2000) seem to imply a more complex picture, in which many distinct cortical regions are involved in shape from motion discriminations. This seems to raise more complex questions, about the bandwidth which would be needed to share spatial information between these cortical regions, if only direct cortico-cortical channels were involved.

In summary, Shape from Motion knowledge sources provide strong reasons for requiring:

- precise representation of positions in all three dimensions
- rapid and precise spatial computations
- parallel operation of many instances of the same type of knowledge source
- collaboration between different knowledge source instances and types, in temporary coalitions to fix the spatial locations of features.
- dynamic linkage between knowledge sources, rather than static links, to form the temporary coalitions

These are important factors in motivating the aggregator hypothesis.

## 9. Geometry First, or Learning First?

The deep learning neural nets which dominate today's vision research are learning-first applications; a neural net needs to learn something before it can do anything.

There is evidence that animal brains do not work this way. There are three types of knowledge source, closely tied to vision, which are pure geometric computations, with no dependence on learning:

- Stereopsis
- Shape from Motion
- Shape from Shading

Everyday experience shows that these work well, even for irregularly shaped objects which are unlearned and unlearnable. They appear to work well with no learning, as in those herd animals that can walk within a few minutes of birth.

They produce a robust 3-D segmentation of any scene, which is a better input for any learning module than a raw visual image. Separate objects are clearly discriminated, and objects are resolved to a invariant scale, at any distance from the viewer.

Therefore, it seems likely that mammalian brains are geometry-first, rather than learning-first; and that learning modules take as their inputs a fully parsed three-dimensional scene, rather than a raw visual image. This seems necessary in order to achieve any learning within biologically realistic numbers of training examples.

The aggregator model is well suited for the geometry-first approach - because it allows the pure geometric knowledge sources to act together to give a robust 3-D segmentation of any scene, even before any learning knowledge source has had time to learn. The input to any learning module is then the set of 3-D positions from the aggregator - with different objects well discriminated and scaled.

There are several current image understanding challenges which test a program's ability to segment and classify objects from large numbers of still images - typically using deep learning neural nets to learn from very large numbers of examples.

If we want these challenges to be biologically more realistic, and to learn from smaller numbers of training examples, then a change of emphasis would be useful. Instead of learning from still images, programs could be required to learn from video sequences (e.g. Cadieu & Olshausen 2012). Then the knowledge sources of shape from motion and shape from shading could be applied to produce robust 3-D segmentations of the scenes, leading to better alignment and scale invariance of training examples, and faster learning. This would lead to safer AI vision systems, which can learn more rapidly about rare or novel features of their environment.

## 10. Conclusions

The dominant current models of spatial cognition are hierarchical Bayesian models of vision, and deep learning neural nets. These have important common features:

1. They are **distributed** models, in which a number of modular learning units (or knowledge sources) cooperate to infer the locations of things in space.
2. They are **hierarchical** models, in that the modules are arranged into a number of layers, with the units of each layer addressing a larger region of space than the connected units in the lower layer
3. Connections between modules are usually **direct** and **permanent**, modelling permanent axonal connections
4. The modules and their direct connections are the main feature of the models; there is usually little else.

The hierarchical modular arrangement of knowledge sources is broadly based on the visual cortex and other areas of the primate cortex. If we include both research into understanding the brain, and industrial work on artificial intelligence, the amount of effort currently devoted to these models is immense.

This paper has argued that as a model of spatial cognition in the mammalian brain, the ingredients (1) - (4) are not sufficient. The fully distributed nature of the models, together with the assumed hard-wired nature of the connections between modules, lead to major problems when scaling to animal performance. I have discussed the problems under the headings:

a) **Scaling of connectivity**: there is a need for temporary coalitions of knowledge sources to determine and track the locations of individual features. Having permanent hard-wired connections between knowledge sources is an inefficient way of providing the connections required for the temporary coalitions. Any other solution implies some other component in the architecture, as well as the knowledge sources.
b) **Feature binding**: Any knowledge source on its own does not have the information required to determine which other knowledge sources it should be bound to, in a temporary coalition to fix the location of some feature. Forming any temporary coalition requires information not present in any knowledge source.
c) **Precise spatial computation and steering**: Most knowledge sources work in terms of small local spatial displacements between features. However, for two or more knowledge sources to collaborate, the representation of any one feature must be correctly spatially aligned in all of them. This requires precise spatial computation of displacements, and precise spatial steering of signals between knowledge sources. Small local knowledge sources in cortex are not well suited for doing this. The need for precise steering implies more wiring costs, if connections are hard-wired.
d) **Fast learning of invariant patterns**: To learn spatial patterns which are invariant under different displacements from the animal, within biologically realistic numbers of training examples, also requires precise spatial steering of features to the learning knowledge sources.
e) **Object Constancy**: Animal brains are finely tuned to exploit a very important Bayesian prior probability, which has been true for all evolutionary time: that for most of the time, most objects do not move. In a fully distributed model, where knowledge sources deal in relative displacements, there is no place where this important constraint can be properly represented.

These considerations point to the need for some extra component in the architecture of the brain.

This paper has proposed that knowledge sources do not communicate information about the spatial locations of features directly to one another; they communicate that information indirectly, through an **aggregator** - which combines the estimates of any feature location from different knowledge sources into an overall best estimate of the position of that feature, in some animal-centred frame of reference.

The aggregator model has benefits over fully-distributed models, under the headings (a) - (e) above:

a) **Scaling of connectivity**: The number of channels needed to communicate spatial information is vastly reduced - approximately from $K^2$ to $K$, where $K$ is the total number of knowledge sources, and is a large number (estimated to be $K = 400$ for a small primate).
b) **Feature Binding**: as the aggregator stores all features by their locations in a common frame of reference, any two features can be spatially bound if they have nearby locations - so they will be transmitted from the aggregator to the same temporary coalitions of knowledge sources. This is the biologically required form of binding; features which overlap in space are likely to be features of the same object.
c) **Spatial computation and steering**: In the aggregator model, there is still a need for precise spatial computation and signal steering. The aggregator can be a specialised large region of the brain devoted to doing it.
d) **Invariant learning of spatial patterns**: The aggregator can provide the precise spatial steering

of sense data and features needed to learn invariant patterns.

e) **Object constancy**: if the aggregator frame of reference is chosen so that most of the time it does not move (stationary objects appear stationary), then the Bayesian prior that 'most objects do not move' can be represented there - and the aggregator can act as a spatial short-term memory.

Therefore, the aggregator model has important advantages over fully distributed models, in meeting these key requirements.

If there is an aggregator, the need to minimise axonal energy consumption implies that the aggregator must be near the physical centre of the brain, and so not in cortex. This and other reasons suggest that the aggregator is the thalamus. Notably, the thalamus is the gateway to cortex for all sense data which carry location information - but not for smells, which carry hardly any location information. The thalamus also has the required reciprocal connections to many parts of cortex.

There is a need to store spatial locations with high precision in the aggregator. This need can be met by a Fourier transform-like representation of space, in which the densities of synapses have a wave-like (sinusoidal) distribution in the brain. With this representation, sigma-pi neurons may be able to compute spatial displacements with high precision, and to carry out the aggregator function of pooling position estimates from different knowledge sources. The thalamus, being extended in all three dimensions, is well-suited to hold this representation.

I propose that the aggregator model is a viable alternative to fully distributed models of spatial cognition, with the potential to solve some of the serious problems facing those models. In many cases, this will not involve abandoning the current distributed models - but will involve adding an aggregator component to them.

In this paper I have described a model of a central aggregator for multiple knowledge sources, applied to the problem of immediate spatial perception. The quantity which is aggregated from multiple knowledge sources, for a purely perceptual cognitive problem, is the negative log likelihood of an object or feature being at a certain location, conditional on sense data. This negative log likelihood is a free energy [Friston 2010]. Through the free energy principle (FEP), different aggregator models could be applied to other cognitive problems in which free energies from a dynamic coalition of knowledge sources need to be summed, such as the problems addressed using active inference [Friston 2013, Friston et al 2017], in which there is value for the animal in satisfying curiosity, or gathering the most useful sense data. Aggregator principles may also be applied to problems of spatial motion planning and control, in a free energy framework.

A model of the role of the pulvinar which has much in common with the aggregator model is the model of [Kanai et al, 2014]. In their model, the pulvinar (the largest nucleus in the primate thalamus, and the nucleus likely to be involved in an aggregator model as a spatial aggregator) has a role in modulating the precisions of cortical knowledge sources, in hierarchical processing with predictive coding. The precision of a knowledge source is the rate at which the gradient of a free energy increases (i.e. a second derivative of the free energy near its minimum), in response to a discrepancy from its predictions (e.g. detected by a source lower in the hierarchy).

Precision is defined in terms of gradients of free energy, which raises the question – what are the gradients with respect to? Which variables? In a gradient $dE/dx$, what is $x$? For spatial cognition tasks, the variables are coordinates in a representation of space. Comparing these gradients is just what the pulvinar does in the aggregator model. To locate an object or feature, it aggregates (sums) free energies from different knowledge sources, to find the location where the free energy is a minimum – that is, where the sum of free energy gradients with respect to any location coordinate is zero.

Thus, the model of Kanai et al. and the aggregator model have much in common, and further investigation of their commonalities would be worthwhile. The two models do not overlap completely, since Kanai et al. use a hierarchical model of cortical processing, and their gradient summations occur in cortex rather than in the thalamus. As I have discussed in the context of the hierarchical models of Olshausen et al., this might require prohibitive numbers of cortico-cortical axons; in the aggregator model, free energies are summed in the thalamus.

Another application of the free energy principle which is close to the model of this paper is the Mathematical Model of Embedded Consciousness by [Rudrauf et al. 2017]. Like this aggregator model, theirs is a model of spatial cognition. Rather than the Euclidean 3-D geometry used in the aggregator model, they use (more powerful) projective geometries, allowing an animal to take different perspectives on its surroundings for different purposes. Spatial perspectives have been outside the scope of this aggregator model, but may be related to the focus of attention discussed in section 4 of this paper. Aggregator models will need to further address issues of attention and perspectives, which may lead to convergence between the models. Interestingly — in free energy formulations of aggregation — optimising the precision of various probabilistic representations is exactly analogous to attentional selection [Parr and Friston 2019] This theme may provide a unifying principle for the functional integration that plays out over several thalamic nuclei; for example, the pulvinar for visual attention and the mediodorsal nucleus of the thalamus in mediating salience [Feldan and Friston 2010; Parr and Friston 2017].

There are also close links to the SAIM model (Selective Attention for Identification) of [Heinke & Humphreys 2015]. While their model is a distributed hierarchical routing model, like those of [Olshausen 1993,1995], its focus on mechanisms of attention and binding addresses many of the issues tackled by the aggregator model, through temporary coalitions of knowledge sources, spatially bound through an aggregator. In the SAIM model, features can be bound (i.e. used by a temporary coalition of knowledge sources) by being part of the same object, as well as by spatial proximity – an area for further exploration in aggregator models. I suggest that for the reasons of scaling, precision, and neural economy described in the early sections of this paper, a SAIM model could use an aggregator-like routing component, in a more hub-and-spoke architecture, that has close connections to the aggregation of free energy above [Abadi et al. 2019]

A model which seems well suited to extend with an aggregator is the model of cortical columns by Hawkins, Ahmad and Cul [2017]. These authors propose that most cortical columns use representations of allocentric (= object-centred) locations; and they say that: "*we don't know how the location signal is generated*". In the hypothesis of this paper, groups of cortical columns are the knowledge sources. Their location signals could come from the aggregator, by thalamo-cortical axons projecting into layer L4; and locations are returned to the aggregator by cortico-thalamic axons from layer L5.

An aggregator component may also be an appropriate addition to the neocortical microcircuit model of [Bennett 2020], which shares several features with Hawkins' models.

There are existing models of cortical function which extend the fully distributed model with an intermediate component which routes signals between cortical modules. Zylberberg et al. (2010) propose a model with a 'router' component, conveying signals between cortical modules. Their model differs from ours in several respects: their router does not perform spatial computations, or aggregate different estimates of the position of a feature; and it appears to be in the cortex rather than the thalamus.

The models of Olshausen et al. (1993, 1995) use direct selective routing of signals between cortical modules, rather than a separate routing component.

To recap how the model of this paper differs from leading hierarchical Bayesian models of spatial cognitions:

At Marr's (1980) Computational level 1, those models use sets of Bayesian inference units cooperating in a hierarchy, to infer the locations of objects in space. At Marr's [1982] Neural Implementation level 3, hierarchical Bayesian models are typically mapped onto brains as in the table below. While the table does not include Marr's level two, this paper has made preliminary proposals for how his level two (of domain-specific data structures and algorithms – in this case for three-dimensional geometry) is addressed.

| Model Component (level 1) | Neural Realisation (level 3) |
|---|---|
| Bayesian inference unit | Cortical module |
| Links between inference units | Cortico-cortical axons |

This paper proposes that the Bayesian hierarchy must be a **dynamic hierarchy**, with dynamic temporary coalitions of inference units. The static mapping in the table above is ill-suited for this, in two main respects:

- Scaling of connectivity costs
- Binding of features to inference units

The dynamic hierarchy is determined by **spatial geometry**: two inference units can share a feature only if it might occur at the same spatial location for both of them. A new computational and linking component, the **aggregator**, is proposed to meet these requirements.